\documentclass[3p,times,nopreprintline]{elsarticle}
\usepackage{amssymb}

\usepackage{lineno}

\usepackage[figuresright]{rotating}

\begin{document}
\begin{frontmatter}

\centering{23rd Conference on Application of Accelerators in Research and Industry, CAARI 2014, San Antonio, TX}
\title{A New Proton CT Scanner}

\author[niu]{\corref{corr_su}S.~A.~Uzunyan}
\author[niu]{G.~Blazey}
\author[niu]{S.~Boi}
\author[niu]{G.~Coutrakon}
\author[niu]{A.~Dyshkant}
\author[niu]{B.~Erdelyi}
\author[niu]{D.~Hedin}
\author[niu]{E.~Johnson}
\author[niu]{J.~Krider}
\author[niu]{V.~Rykalin}
\author[niu]{V.~Zutshi}
\author[fnal]{R.~Fordt}
\author[fnal]{G.~Sellberg}
\author[fnal]{J~.E.~Rauch}
\author[fnal]{M.~Roman}
\author[fnal]{P.~Rubinov}
\author[fnal]{P.~Wilson}
\author[delhi]{M.~Naimuddin}
\address[niu]{Department of Physics, Northern Illinois University, DeKalb, IL 60115, USA}
\address[fnal]{Fermi National Accelerator Laboratory, Batavia, IL 60510, USA}
\address[delhi]{Delhi University, 110007, India}
\cortext[corr_su]{Corresponding author. E-mail address: serguei@nicadd.niu.edu}
\begin{abstract}
%
The design, construction and results of preliminary testing of a 2$^{nd}$ generation proton CT scanner are presented. All current treatment planning systems at proton therapy centers use X-ray CT as the primary imaging modality for treatment planning to calculate doses to tumor and healthy tissues. One of the limitations of X-ray CT, when used for proton treatment planning, is the systematic error in the conversion of X-ray attenuation coefficients to relative (proton) stopping power, or RSP. This results in more proton range uncertainty, larger target volumes, and, therefore, more dose to healthy tissues. To help improve this, we built a novel scanner capable of high dose rates, up to 2~MHz, and large area coverage, 20~x~24~cm$^2$, for imaging an adult head phantom and reconstructing more accurate RSP values.
\end{abstract}

\begin{keyword}
proton Computed Tomography \sep range stack detector \sep fiber tracker
\end{keyword}
\end{frontmatter}
%
\section{Introduction\label{intro}}
Northern Illinois University in collaboration with Fermi National Accelerator Laboratory (FNAL) and Delhi University has been designing and building a proton CT scanner \cite{bruges_paper} for applications in proton treatment planning. In proton therapy, the current treatment planning systems are based on X-ray CT images that have intrinsic limitations in terms of dose accuracy to tumor volumes and nearby critical structures. Proton CT aims to overcome these limitations by determining more accurate relative proton stopping powers directly as a result of imaging with protons. At present, the proton RSPs for various tissues, as derived from X-ray CT, produce range uncertainties \cite{xct_uncert} of about 3 to 4\%. We hope to reduce this to approximately 1\% of the total range using proton CT. In addition, doses potentially much lower than those from X-ray CT are possible and absence of artifacts from high density dental or other implants will add to higher quality images. Proton CT image reconstruction is best performed by reconstruction of the individual proton tracks and measurement of their energy losses in the scanned volume. The number of protons to acquire for a typical head-size volume scan is of the order of one billion. To finish the scan in a time acceptable for medical use, the track collection rate should be of the order of 2~MHz, which requires fast tracker and energy detectors. To date, two proton CT scanners are under development in the United States. A system that uses silicon strip technology for the tracker planes and five plastic scintillators for the range measurements was built at the Santa Cruz Institute of Particle Physics and is undergoing testing at Loma Linda University Medical Center~\cite{ll_scaner}. We describe the proton CT scanner that is based on a fiber tracker and a scintillator-stack range detector, which has been developed at Northern Illinois University in conjunction with FNAL in Batavia, IL.  
\begin{figure}[ht]
\centering
\includegraphics[scale=.40]{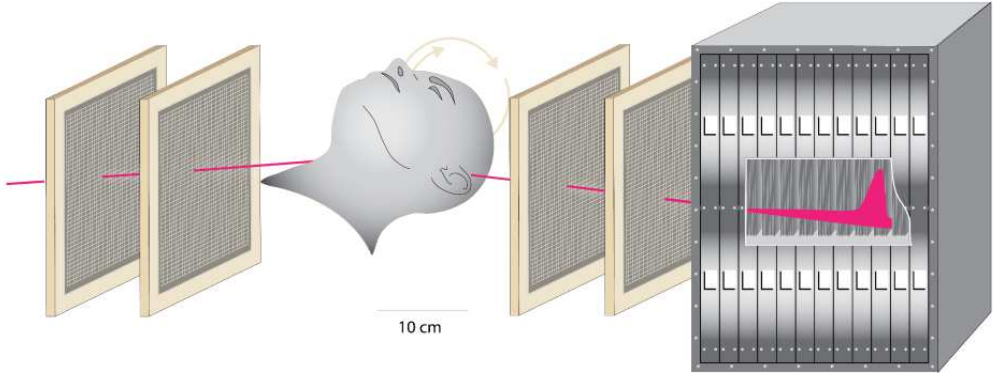}
\caption{\label{fig:pct_design} Four (X,Y) stations measure the proton trajectory before and after the patient.
A  stack of 3.2~mm thick scintillator tiles is used to measure the residual energy or range after the patient.
}
\end{figure}
Fig.~\ref{fig:pct_design} shows a schematic of the proton CT scanner. It consists of eight planes of tracking detectors, providing two X and two Y coordinate measurements before and after the patient, respectively. This provides the information for finding the individual proton trajectory through the head to correct, as much as possible, for multiple Coulomb scattering in the patient. A “most likely path” formalism~\cite{mlp_form} is used to find which voxels, which are of the order of 1~mm$^3$, are crossed by every track. In addition, a calorimeter consisting of a stack of thin scintillator tiles is used to determine the water equivalent path length (WEPL) of each track through the head. The X-Y coordinates and WEPL are required input for image reconstruction software to find RSP values of each voxel in the head and to generate the corresponding 3D image.
\section{Design specifications}
In addition to a high data rate of 2~MHz, we wish to cover a large enough area to image an adult human head so that table motion is not required or that we do not need to splice data from multiple scans to make an image long enough along the body axis. For head scans, we have chosen a maximum head size of 23~cm diameter and a length along the body axis of 20~cm. This will allow imaging from the top of the head down to the jaw bone in one 360$^\circ$ gantry rotation. An incident proton beam energy of 200~MeV with a range of 26~cm in water can be used for head-size imaging. This proton CT detector is compatible with the geometric constraints of most proton treatment nozzles and patient positioners. Beam spreading from an effective source in the nozzle sets the detector sizes required for cone beam geometry. Keeping multiple coulomb scattering in the tracking detectors manageble requires a reduction of the mass of the detectors. For this reason, each tracking plane has a water equivalent thickness (WET) less than 1~mm.
\section{Detector Design and Construction\label{front-end}}
\begin{figure}[ht]
\centering
\includegraphics[scale=0.45]{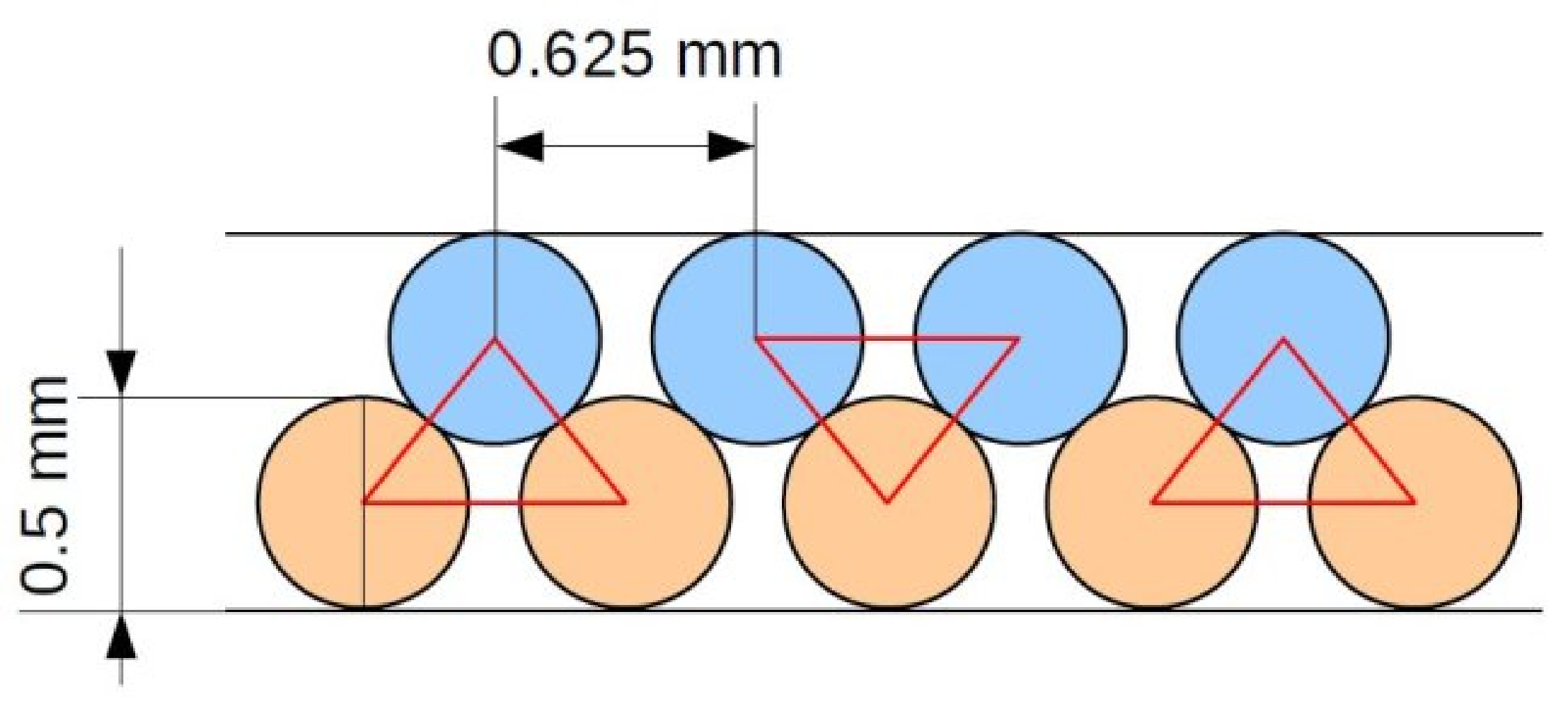}
\caption{\label{fig:fiber_conf} The double layer configuration of fibers which are bundled into 
triangular triplets for readout through a single SiPM. Spacing between bundles is 0.94~mm.}
\end{figure}

\begin{figure}[ht]
\centering
\includegraphics[scale=.95]{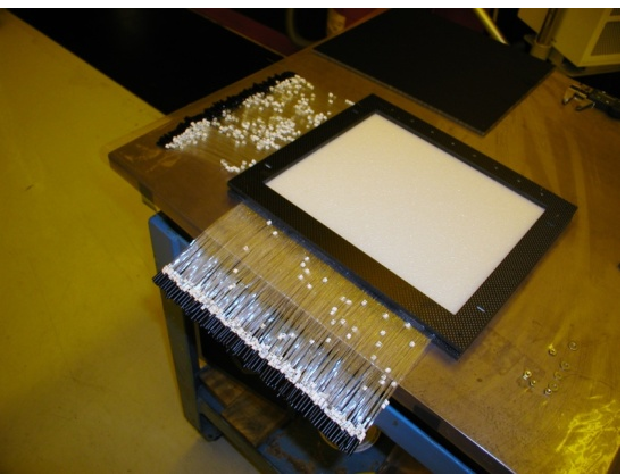}
\includegraphics[height= 4.5 cm]{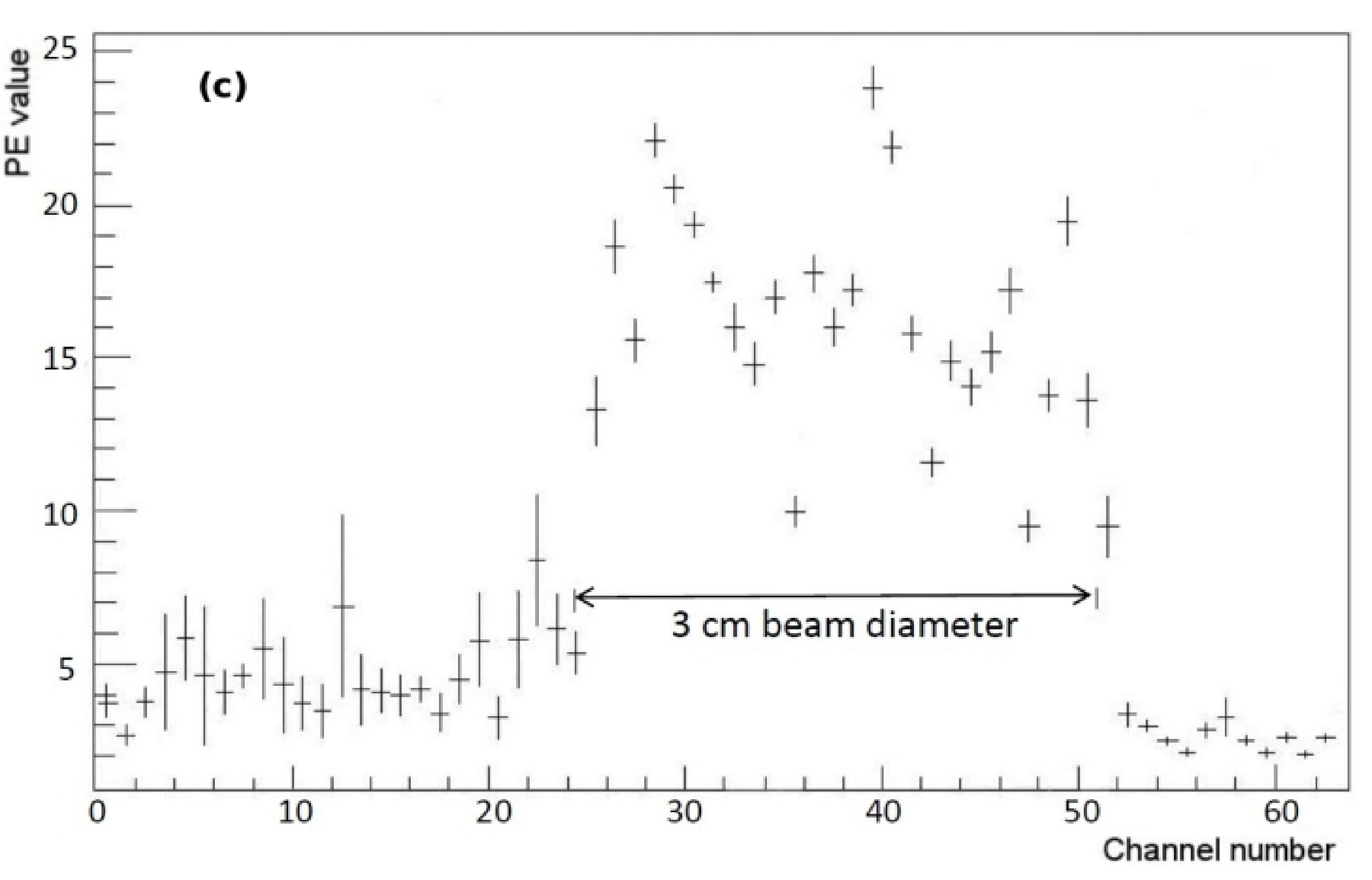}
\leftline{ \hspace{4.2cm}{\bf (a)} \hfill\hspace{1.6cm} {\bf (b)} \hfill}
\caption{\label{fig:tracker} (a) An (X,Y) fiber tracker station assembled on 2~mm ROHACELL\textsuperscript{\textregistered} substrate. 
 (b) The average photo-electron counts per fiber tracker bundle for 200~MeV protons measured with a fiber tracker plane prototype.}
\end{figure}

 In order to have low-mass detectors with high proton rates and continuous area coverage over a large area, the tracker was constructed from 0.5~mm diameter polystyrene scintillating fibers from Kuraray~\cite{Kuraray}. Fibers were initially cut to 50~cm length, then laid flat and doubled layered (see Fig.~\ref{fig:fiber_conf}) on a low density, 0.03~g/{cm$^3$}, 2~mm thick ROHACELL\textsuperscript{\textregistered}  substrate with machined grooves, and glued to hold the fibers in place with close spacing to avoid gaps in detecting passing protons. The entire assembly is supported by carbon fiber frames. A photograph of one tracker plane, 20~x~24~cm, is shown in Fig~\ref{fig:tracker}a.
 
 Fibers are grouped in triplets, called bundles, according to Fig.~\ref{fig:fiber_conf} which give a pitch between bundles of 0.94~mm. Each bundle is read out into silicon photo multipliers (SiPMs), produced by CPTA~\cite{cpta151} which are mounted on Techtron blocks that connect each of them to a fiber triplet. The SiPMs chosen have the best chromatic (or wavelength) match to the Kuraray scintillators. One end of each fiber is polished and mirrored. The other end is polished and mechanically pressed to a SiPM. To measure signal to noise ratio of the fiber bundles, a fiber tracker plane prototype was also exposed to a 200~MeV proton beam. The result is shown in Fig.~\ref{fig:tracker}b, demonstrating 10 to 25 photo-electrons per proton per channel in the beam spot area. The rms spatial resolution of each tracker plane is given by the pitch divided by $\sqrt{12}$, or 0.27~mm. The integrated WET of each tracker along the beam direction is less than 1~mm. With four planes of 20~x~24~cm$^2$ in area and four planes with 24~x~30~cm$^2$ in area, there are about 2100 channels of readout for the entire tracker. 
 
 
 %
 The calorimeter chosen for this design is a proton range detector which consists of a stack of 96, 3.2~mm thick, polyvinyltoluene (PVT) scintillating tiles, with 0.006~mm aluminized mylar between adjacent tiles. Each tile, 27~x~36~cm$^2$ in area, is machine grooved to embed a 1.2~mm diameter wavelength shifting (WLS) fiber that weaves four times across the tile for improved light collection efficiency. Both ends of the WLS fiber are read out through SiPMs. This requires 192 channels of readout for the calorimeter. Each SiPM signal is amplified and digitized for later analysis for fitting to the shape of a Bragg peak to determine the proton range in the calorimeter. Water equivalent blocks can be used to calibrate range measured in the calorimeter~\cite{WEPL_Calibration}.
 
 An intrinsic limitation in any proton calorimeter is the range (or energy) straggling due to the combined mass represented by the patient plus calorimeter. In near water equivalent materials such as brain tissue and PVT scintillator, the sum of energy straggling in the human head and calorimeter for the 200 MeV beam is almost constant and approximately equal to $\pm$~mm~\cite{janni_tables}. Therefore, there is little incentive to produce tiles less than 3~mm thickness.

\begin{figure}[ht]
\centering
\includegraphics[height= 3.4cm]{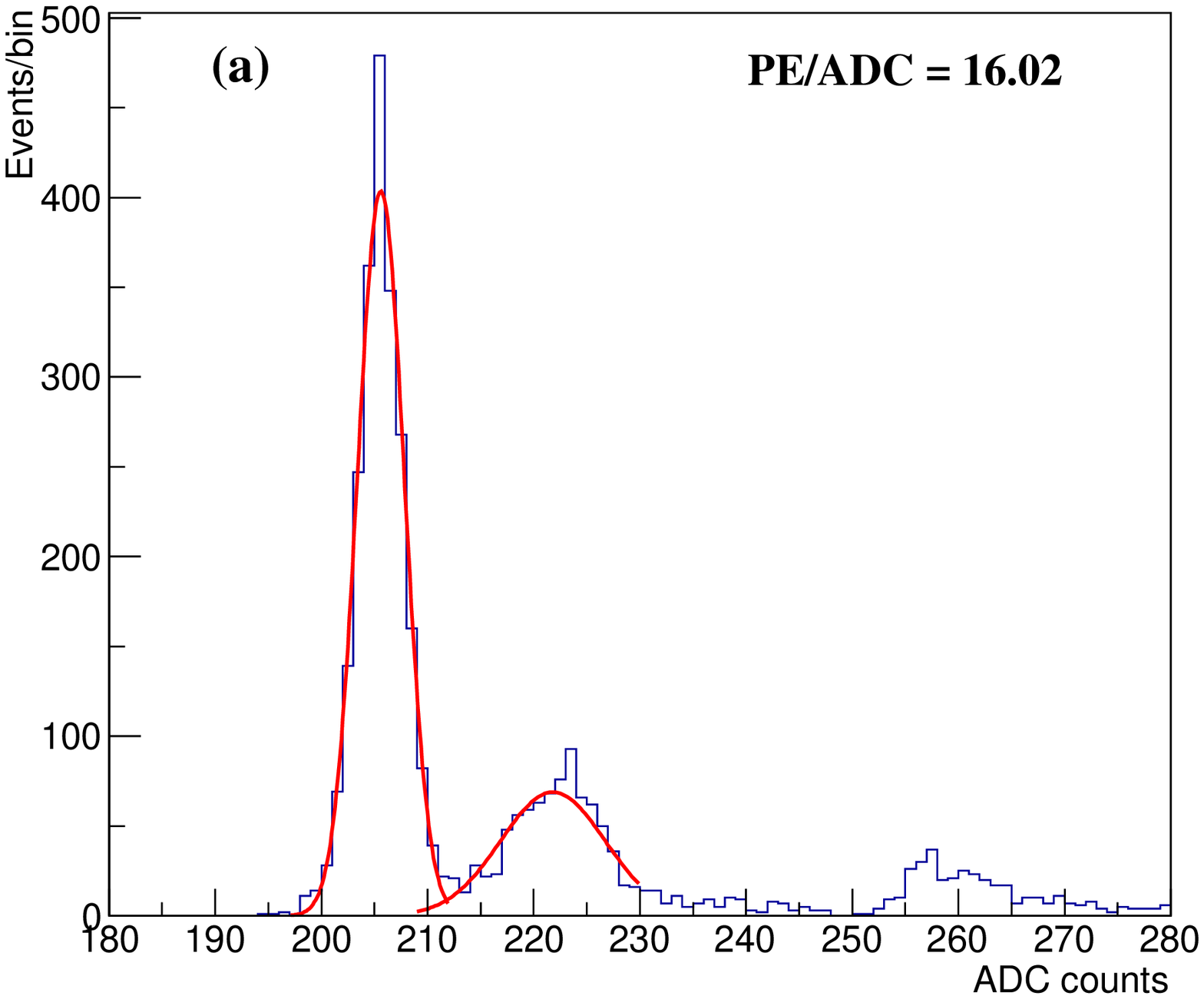}
\includegraphics[height= 3.38cm]{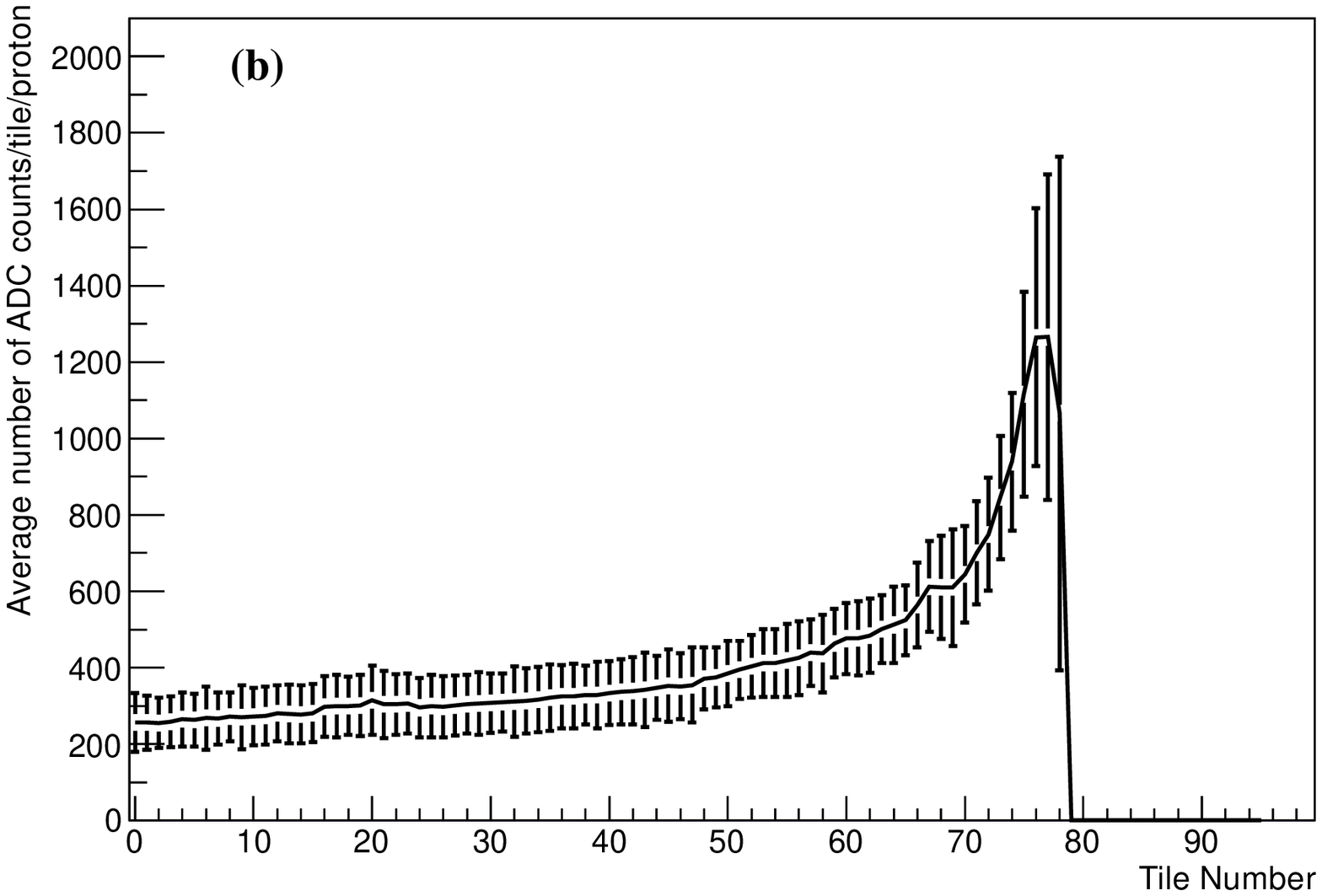}
\includegraphics[height= 3.38cm]{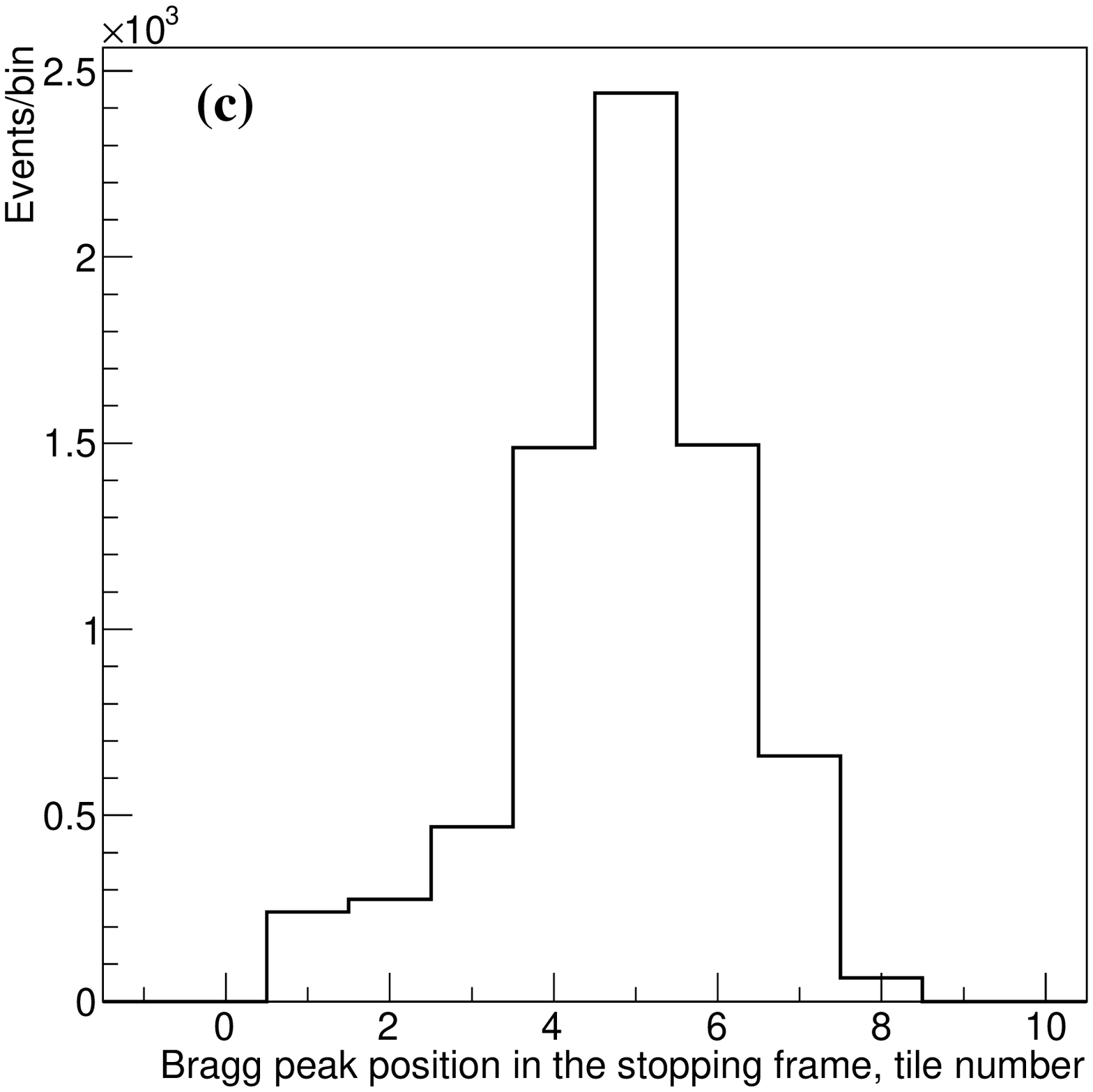}
\caption{\label{fig:bragg_peak}
(a) Pulse height spectrum in ADCs showing the  pedestal (first peak) and the single photo-electron noise signal (second) peak for a single calorimeter tile. 
(b) The  ADC distribution as a function of tile number for 200~MeV protons. The error bars represent $\pm$1~RMS about the average. 
(c) The distribution of the Bragg peak position (the tile number with the maximum of the signal in the stopping calorimeter frame) for 200 MeV protons. A Gaussian fit of this distribution gives the position resolution of 3.7~mm.}
\end{figure}

 The 96 tile calorimeter was built and underwent first tests with a 200~MeV proton beam at Central DuPage Hospital in Warrenville, IL. Examples of a pedestal distribution and a single photoelectron distribution from a calorimeter tile are shown in Fig.~\ref{fig:bragg_peak}a. Figure~\ref{fig:bragg_peak}b shows the Bragg peak from a sample of 200 MeV protons. The distribution of the Bragg peak position (the tile number with the maximum of the signal in the stopping calorimeter frame) for this sample is shown in Fig.~\ref{fig:bragg_peak}c.
\section{Electronics \label{pade}}

\begin{figure}[ht]
\centering
\includegraphics[scale=.46]{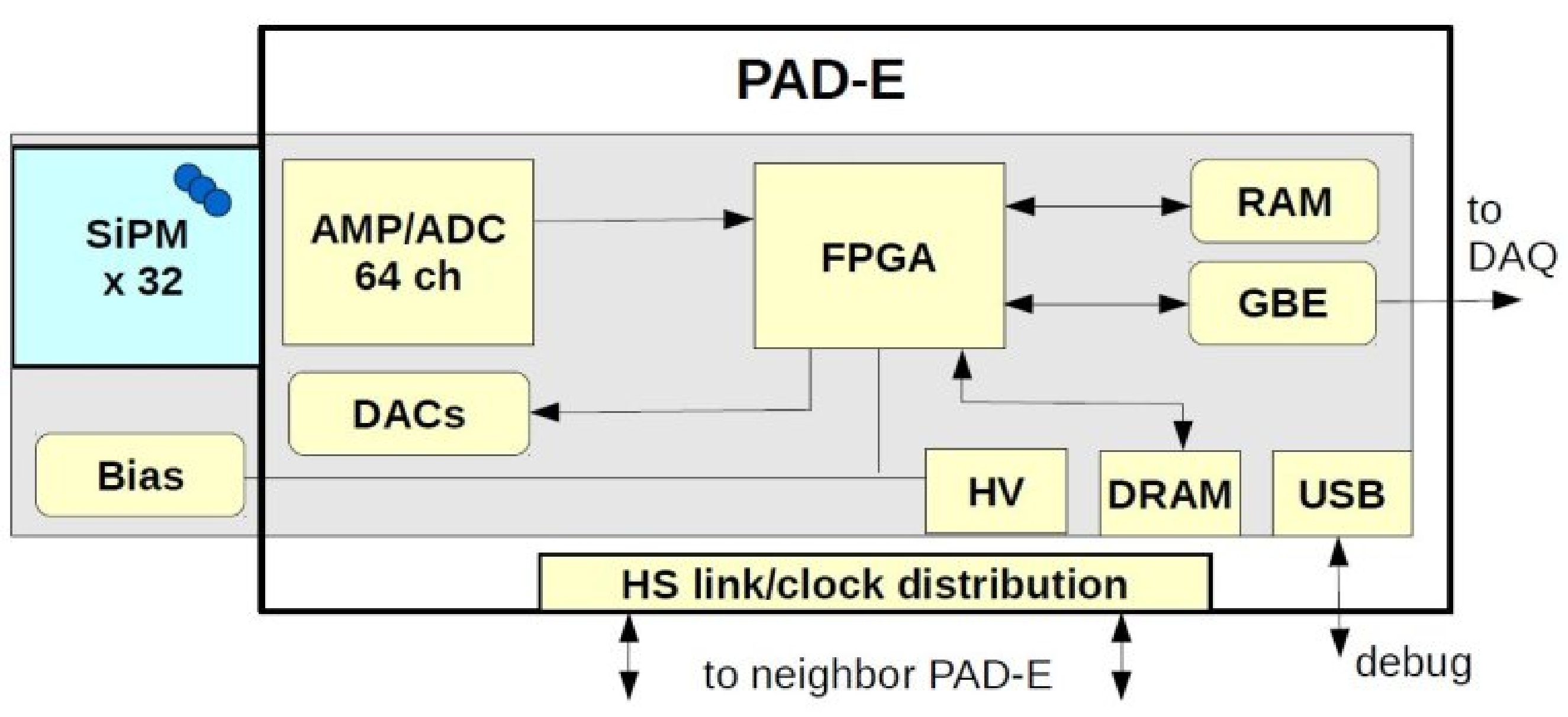}
\caption{\label{fig:pct_pade}A diagram of the front end electronics (PAD-E) for amplifying, 
digitizing and storing data before shipping to the DAQ at one millisecond intervals.}
\end{figure}

The electronics that reads out the SiPMs consists of a custom board with preamplifiers, digitizers, and ethernet readout (PAD-E, see Fig.~\ref{fig:pct_pade}). This custom board uses COTS components to provide readout for up to 32~SiPM channels in a 220~mm~x~100~mm format that fits into a standard 3U sub-rack. The same board is used for readout of the trackers and the calorimeter. The digitization of the signals from SiPMs, after appropriate amplification and shaping, is 12 bits per channel at 75~MSPS. The PAD-E is completely self contained and generates the bias for the SiPMs (one bulk voltage, but with a 3~V adjustment range for each SiPM). It also contains an FPGA for processing of all of the data generated by the SiPMs, memory for buffering up to 128~MB of data, and a gigabit Ethernet interface for pushing data directly to the data acquisition (DAQ) system. Other support circuitry includes temperature sensors for the SiPMs, clock management, and a high-speed USB port for debugging. Parameters such as the board’s Ethernet address or the correct bias voltage for the SiPMs are stored in a small flash memory on the board. The PAD-E is powered by a single 5~V power supply and has a power consumption of up to 15~W for 32~SiPM electronics channels. Each board locks to the clock provided by one board in each of the 9 sub-racks. Each of these boards, in turn, locks to a ``master'' board, which has a free running crystal clock. Run control is accomplished by communicating with this ``master'' over Ethernet. 

The scanner is ``self-triggered'' in the sense that any channel with a signal above threshold will be time-stamped and stored in a local buffer for readout. A synchronous signal allows all boards to provide a timestamp that is used by the DAQ system to associate the data from different parts of the detector for a single proton history. Data from signals in the detector is highly compressed (only fiber address and timestamp from the trackers, compressed amplitude and time stamp from the calorimeter) and sent to the DAQ as soon as it is available. A synchronization signal which circulates across all boards approximately once per millisecond initiates a packet or “frame” of data readout from PAD-E memory to DAQ memory via 1~Gbit/s ethernet with only slight dead time penalty. A ``footer'' with error messages can be sent with each packet as well. Organizing the data into these one millisecond ``time frames'' allows for a relatively small timestamp (16~bits of 75~MHz clock cycles) and allows the DAQ to monitor the integrity of the data.
\section{Data Acquisition System \label{daq}}
\begin{figure}[ht]
\centering
\includegraphics[scale=1.80]{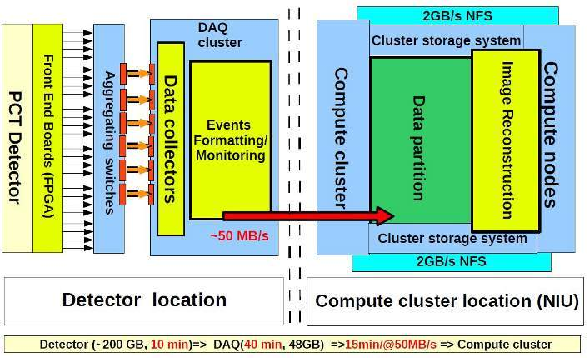}
\caption{\label{fig:pct_daq} DAQ computer cluster (left) and GPU cluster computer (right) 
for image reconstruction. Data collectors hold hit locations and ADC amplitudes of the scintillator tiles 
with a time stamp for coalescing into tracks during post processing. 
Events are then sent to the image reconstruction computer 
cluster shown on the right side of the figure. 
}
\end{figure}

The structure of the DAQ system~\cite{su_daq} for the pCT scanner is shown on the left side of Fig.~\ref{fig:pct_daq}. The frontend electronics sends data to the DAQ via 1~Gbit/s Ethernet lines using UDP protocol.  Each frontend event contains timestamped data from 8 tracker planes and the 96-tile scintillator stack. We calculated that each event will generate about 25 bytes from the 8 hits on the 8 planes and about 75 bytes from the 96 tiles. For a 10 minute scan with 90 projection angles at a data rate of 2 million protons per second, we expect 200~MB/s written to RAM by 24 data collectors running on six interconnected Linux workstations. At the end of the scan, the back-end DAQ will write data to disk and subsequently, through post processing of the data, build the proton candidate events, formed from tracker and scintillator hits within 100~ns time window.  These events are further analyzed for presence of pile-up (hits from same stack frames or excess of  hits in the tracker)  to construct only  the single proton candidates and to extract single proton histories in the format for image reconstruction, i.e., 4~X and 4~Y coordinates, WEPL, and beam (or phantom) rotation angle. 
\section{SUMMARY}
\begin{figure}[ht]
\centering
\includegraphics[scale=1.90]{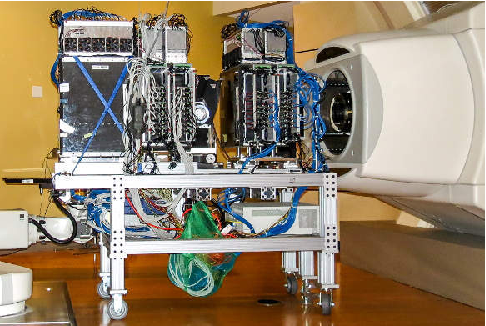}
\caption{\label{fig:assembled_scaner} Fully assembled proton CT scanner at CDH Proton center. 
From right to left, beam enters upstream tracker planes followed by downstream 
tracker planes and finally the calorimeter. The gap in the middle is where the rotation 
stage for rotating the head phantom in the horizontal plane is placed.}
\end{figure}

The NIU Phase~II proton CT scanner is fully assembled and installed for tests in the 200~MeV proton beam at Central DuPage Hospital (CDH) in Warrenville, IL, USA. Fig.~\ref{fig:assembled_scaner} shows the scanner mounted on a cart in the treatment room. After system commissioning, a CIRS head phantom~\cite{cirs_norfolk} will be inserted between tracker planes to collect data for image reconstruction on a CPU/GPU compute cluster~\cite{duffin_gpu}. This compute cluster has been tested with data acquired with an earlier prototype scanner  \cite{caari_2012,phaseI_pct}, and has demonstrated high quality 3D image reconstruction of the 14~cm diameter, spherical Lucy\textsuperscript{\textregistered} 3D QA phantom from Standard Imaging, Inc.~\cite{phaseI_pct_II}. The detailed project documentation can be found at~\cite{niu_pct_web}.
%
\section{Acknowledgements}
We wish to thank Central DuPage Hospital and Loma Linda University Medical Center 
for the generous use of their facilities.
This project was supported by funds from the Department of Defense.


\bibliographystyle{elsarticle-num}







\end{document}